# Statistical inference as Green's functions

**Hyun Keun Lee[1], Chulan Kwon[2], and Yong Woon Kim[3]**

[1] Sungkyunkwan University, Department of Physics, Suwon 16419, Korea.

[2] Myongji University, Department of Physics, Yongin 17058, Korea.

[3] Korea Advanced Institute of Science and Technology, Department of Physics, Daejeon 34141, Korea

## Abstract

Statistical inference from data is a foundational task in science. Recently, it has received growing attention for its central role in inference systems of primary interest in data sciences and machine learning. However, the understanding of statistical inference is not that solid while remains as a matter of subjective belief or as the routine procedures once claimed objective. We here show that there is an objective description of statistical inference for long sequence of exchangeable binary random variables, the prototypal stochasticity in theories and applications. A linear differential equation is derived from the identity known as de Finetti's representation theorem, and it turns out that statistical inference is given by the Green's functions. Our finding is an answer to the normative issue of science that pursues the objectivity based on data, and its significance will be far-reaching in most pure and applied fields.

## Introduction

A prototype of statistical inference is arguing the probabilistic feature of binary outcome with previous observations. The related studies date back to the seminal works by Bayes [1] and Laplace [2], which was ahead of the times and still remains a significant issue to date. Consider a black box that randomly assigns a binary number 0 or 1 in each trial, and suppose that 1 is observed $n$ times out of $N$ trials. In this situation, one may seek the conditional probability $f(\theta|D)$, where $\theta$ is the probability of 1 in a trial and $D$ is a short notation for the observation data. Intuitively, a probability density localized around $\theta = \sigma_N \equiv n/N$ could be a candidate. As the localization width is expected to decrease for increasing $N$, it is supposed that

$$\lim_{N\to\infty} f(\theta|D) = \delta(\theta - \sigma_\infty) \cdots (1)$$

for $\sigma_\infty \equiv \lim_{N\to\infty} \sigma_N$, where $\delta(x)$ is Dirac's delta function [3]. This limiting property of the data-based inference is a spirit of science [1,2,4-6].

## Two streams of statistical inference

However, the justification of Eq. (1) remains rather unsatisfactory. What Eq. (1) is was examined for the first time in Bayesianism [1,2,7-11], and the attempt arrived at a claim that it must be a belief. According to Bayesianism, any $f(\theta|D)$ is possible in belief, and then the limiting behavior in Eq. (1) becomes a matter of subjective choice of $f(\theta|D)$ therein. It is unnatural to consider a scientific meaning or justification for the sentence/equation including such $f(\theta|D)$ that is provided as a belief.

Meanwhile, frequentism [12-15] is the other stream, positioned against the subjectivity. In frequentism, for an objective stance, the entity $f(\theta|D)$ is rejected because of apparently no objective way to quantify it. As a consequence, there remains no room to play a role in the situation where $f(\theta|D)$ is invoked like Eq. (1). Still, there are wide (mis)uses [3,16-20] of frequentism in most science and engineering fields while expecting objective $f(\theta|D)$ or similar one in routine procedures. This customary but improper practice seems uncontrollable because the unintentional misuse of $f(\theta|D)$ in frequentism that was once claimed objective [21-25] is quite attractive. Such uses are the circumstantial evidences that the notion $f(\theta|D)$ is very natural and intuitive in the situation of statistical inference.

## Goal of this work: discerning $f(\theta|D)$ from belief

As the central entity in statistical inference, $f(\theta|D)$ should be treated in a very reliable way. However, the apparent intractability of $f(\theta|D)$ is regarded as the reflection of the claimed subjectivity in one stream, and this provokes the rejection of $f(\theta|D)$ in the other stream to keep an objective stance of science. Inference studies by subjective quantification or those in the absence of the entity $f(\theta|D)$ cannot make a scientifically meaningful progress. Thus if the intractability of $f(\theta|D)$ is resolved, the nonconstructive situation will drastically change.

In this work, we will demonstrate that $f(\theta|D)$ can be specified in objective way for long sequence $D$ of exchangeable binary random variables [26], for which $f(\theta|D)$ was studied for the first time [1,2] and also which are still significant in modern theories and applications [26-34]. We find out that $f(\theta|D)$ is given by the Green's functions [3] of a differential equation derived from de Finetti's representation theorem [26]. The $f(\theta|D)$ obtained this way justifies Eq. (1) without additional argue or implementation.

## Results

A random sequence $(x_1, x_2, ...)$ is said to be *exchangeable* when its probability $P(x_1, x_2, .., x_N)$ has permutation symmetry such as $P(x_1, x_2, .., x_N) = P(x_{\pi(1)}, x_{\pi(2)}, .., x_{\pi(N)})$ for all $N$ and permutations $\pi$. Let $x_i$ be binary with $x_i = 0$ or $1$, and $D = (x_1, x_2, .., x_N)$ be the $N$-digit exchangeable sequence. De Finetti's representation theorem [26] states that any exchangeable random sequence is a mixture of independent and identically distributed random variables: $P(D) = \int d\theta \prod_{i=1}^{N} \theta^{x_i}(1-\theta)^{1-x_i} f(\theta)$ for a mixture weight $f(\theta)$. In the following, this identity will be systematically converted into an expression stating what $f(\theta|D)$ is.

In order to efficiently exploit the exchangeability, we introduce the random variable $\sigma_N \equiv \sum_{i=1}^{N} x_i /N$, the relative frequency of 1 in $D$, and its probability $P_{\sigma_N}$. By the exchangeability, we have $P_{\sigma_N} = \binom{N}{N\sigma_N} P(D)$ for the binomial coefficient $\binom{m}{n}$. We shall derive a differential equation for $f(\theta)$, and then use it to specify $f(\theta|D)$. For this, we first inspect the Laplace transform [3] of $P_{\sigma_N}$, defined as $\sum_{\sigma_N} e^{-s\sigma_N} P_{\sigma_N}$.

### De Finetti's relation with finite-$N$ correction

Expanding $\sum_{\sigma_N} e^{-s\sigma_N} P_{\sigma_N}$ in powers of $1/N$, and then using the inverse Laplace transformation [3], we find up to the leading order of $1/N$ (Methods I-V) that

$$\hat{L}_\theta f(\theta) \equiv \left(1 + \frac{1}{2N}\frac{d^2}{d\theta^2}\theta(1-\theta)\right)f(\theta) = \sum_{\sigma_N} P_{\sigma_N}\delta(\theta - \sigma_N). \cdots (2)$$

Equation (2) is an inhomogeneous differential equation for $f(\theta)$ with heterogeneity arising from delta-function peaks weighed by $P_{\sigma_N}$.

A foundational issue in the study of statistical inference is the relation between probability $\theta$ and frequency $\sigma_N$, which actually led to the birth of Bayesianism [1]. Later, de Finetti [26] showed that in the limit of infinite $N$, $\int_0^\theta d\theta' f(\theta') = \lim_{N\to\infty} \sum_{\sigma_N \leq \theta} P_{\sigma_N}$. The finite-$N$ relation is, however, unknown yet [11]. The first achievement of this work is Eq. (2) where the leading $\mathcal{O}(1/N)$ correction is revealed. One may integrate Eq. (2) to obtain $\int_0^\theta d\theta' f(\theta') + \frac{1}{2N}\frac{d}{d\theta}\theta(1-\theta)f(\theta)\Big|_{\theta^+} = \sum_{\sigma_N \leq \theta} P_{\sigma_N}$, which shows how de Finetti's relation is modified with the leading correction owing to finite $N$. The statistical inference we will report below follows from Eq. (2).

**Statistical inference via Green's functions**

We denote a Green's function of the linear differential operator $\hat{L}_\theta$ by $G_{\sigma_N}(\theta)$, which satisfies $\hat{L}_\theta G_{\sigma_N}(\theta) = \delta(\theta - \sigma_N)$. Then, the formal solution of Eq. (2) is given by $f(\theta) = \sum_{\sigma_N} P_{\sigma_N} G_{\sigma_N}(\theta)$, which can be rewritten as $f(\theta) = \sum_D P(D) G_{\sigma_N}(\theta)$ with $P_{\sigma_N} = \binom{N}{N\sigma_N} P(D)$. On the other hand, it is straightforward to write $f(\theta) = \sum_D f(\theta|D)P(D)$. Comparing these two equations for $f(\theta)$ yields

$$f(\theta|D) = G_{\sigma_N}(\theta). \cdots (3)$$

This remarkable finding implies that the intractability of $f(\theta|D)$ can be resolved and also that statistical inference is a task of finding the associated Green's functions instead of choosing a belief.

We note that $\hat{L}_\theta$ is a hypergeometric differential operator [3] (Methods V), whose homogeneous solution is given by the linear combination of the hypergeometric functions $\mathcal{H}(\theta) = {}_2F_1(c_+, c_-, 2; \theta)$ for $c_\pm = (3 \pm \sqrt{8N+1})/2$ and $\mathcal{R}(\theta) = \mathcal{H}(1-\theta)$. For a Green's function of the linear differential operator $\hat{L}_\theta$, one may try $g_{\sigma_N,d}(\theta) = h_d \mathcal{H}(\theta) + r_d \mathcal{R}(\theta) > 0$ in the interval $\sigma_N < \theta < d/N$ (or $d/N < \theta < \sigma_N$), which vanishes elsewhere. The prefactors $h_d$ and $r_d$ are chosen from the condition $\hat{L}_\theta g_{\sigma_N,d}(\theta) = \delta(\theta - \sigma_N)$, so that $g_{\sigma_N,d}(\theta)$ becomes an instance of $G_{\sigma_N}(\theta)$. See Methods VI for more technical details of $g_{\sigma_N,d}(\theta)$. An example is shown in Fig. 1(A).

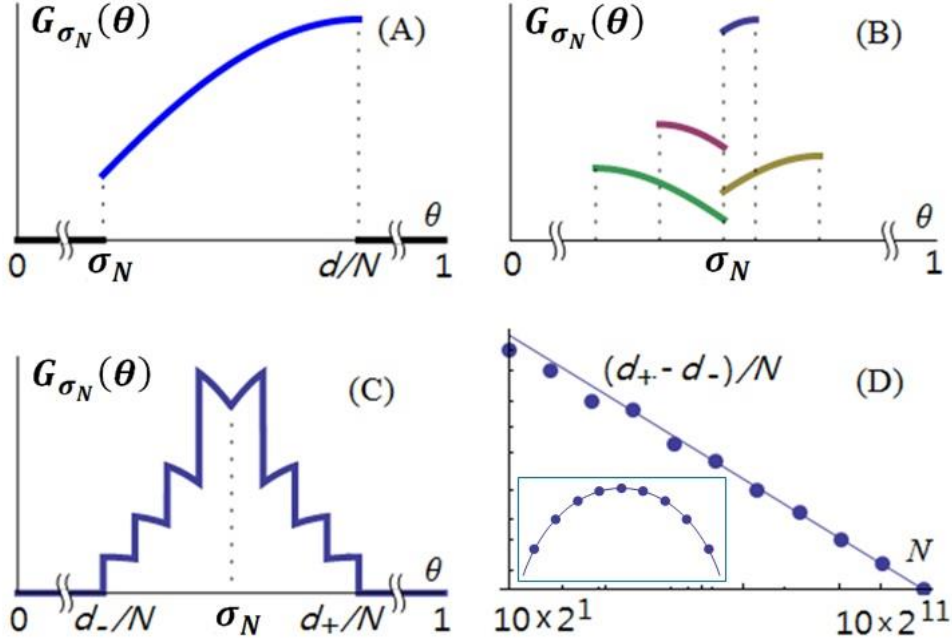

**Fig. 1**: **Illustration of a few $G_{\sigma_N}(\theta)$ and its localization width.** The function $G_{\sigma_N}(\theta)$ for $\sigma_N = 0.5$ and $N = 80$ is shown in (A), which is given by $g_{\sigma_N,d}(\theta)$ for $d = 44$ [see text and Methods VI]. More cases with $d = 41, 38, 43$, and $36$ are presented in (B) (from top to bottom). A general case of a linear superposition of $g_{\sigma_N,d}(\theta)$ with equal weights, that is, $G_{\sigma_N}(\theta) = \sum_d g_{\sigma_N,d}(\theta)/8$ for $d = 40 \pm 1,2,3,4$ is given in (C). In (A-C), the vertical axes are in their own units for visualization. In (D), $(d_+ - d_-)/N$, the width of $G_{\sigma_N}(\theta)$, is displayed in log-log scale, where the guiding line is $1.1/\sqrt{N}$ and the data points are obtained for $N = 10 \times 2^m$ with $m = 1, 2, \ldots, 11$. The inset shows the data points $\delta \equiv (d_+ - d_-)/(2\kappa\sqrt{N})$ from $\sigma_N = 0.1, 0.2, \ldots, 0.9$ for large $N$ on the $\sigma_N$-$\delta$ plane in linear-linear scale, where the guiding curve is $\sqrt{\sigma_N(1 - \sigma_N)}$ for $\kappa = 1.1$ and the axes are omitted for simplicity.

For a given $\sigma_N$, various values of $d$ are possible as shown in Fig. 1(B). There exist such $d_\pm = d_\pm(\sigma_N)$ that impose bounds on $d$, i.e., $d_- \leq d \leq d_+$. Then, the general construction of $G_{\sigma_N}(\theta)$ reads

$$G_{\sigma_N}(\theta) = \sum_{d=d_-}^{d_+} w_d g_{\sigma_N,d}(\theta) \cdots (4)$$

for weights $w_d \geq 0$ with normalization condition $\sum_d w_d = 1$. Various configurations of $\{w_d\}$ manifest the freedom of $G_{\sigma_N}(\theta)$. A case of $G_{\sigma_N}(\theta)$ with uniform $w_d$ is shown in Fig. 1(C). We numerically confirm that $d_\pm(\sigma_N) = N\sigma_N \pm \kappa\sqrt{N\sigma_N(1-\sigma_N)}$ with $\kappa = 1.1(1)$ for large $N$ [see Fig. 1(D)]. Thus, $G_{\sigma_N}(\theta)$ is localized in the interval $\left(\sigma_N - \kappa\sqrt{\sigma_N(1-\sigma_N)/N}, \sigma_N + \kappa\sqrt{\sigma_N(1-\sigma_N)/N}\right)$. We remark that nonnegative $G_{\sigma_N}(\theta)$ does not exist for $\sigma_N = 0$ and 1. This exception is practically unproblematic as such $D$ with $\sigma_N = 0$ and 1 is not practical for large $N$.

**Extraneousness of homogenous solutions**

Equation (2) is derived for probability $P_{\sigma_N}$, the nonnegative heterogeneity sources of unit sum. For this reason, any homogeneous solution $h(\theta)$ of $\hat{L}_\theta h(\theta) = 0$ is extraneous. Therefore, no homogeneous solution is considered in constructing $G_{\sigma_N}(\theta)$. The extraneous solutions in the presence of homogeneous solutions are expected to be the mathematical realization of the subjectivity claimed in Bayesianism because of their considerable freedom and profiles seemingly independent of $P_{\sigma_N}$. A detailed discussion of this topic is outside the scope of this article, and shall be separately reported elsewhere.

# Discussion

**Justification of Eq. (1)**

Since $f(\theta|D) = G_{\sigma_N}(\theta)$ for large $N$ [Eq. (3)], $f(\theta|D)$ is localized near $\theta = \sigma_N$ and its width is at most $\approx 2\kappa\sqrt{\sigma_N(1-\sigma_N)/N}$. Although a localization around $\sigma_N$ is intuitive, to argue whether it is substantial or not is highly nontrivial. When referred to Bayesianism of subjective belief on intractable $f(\theta|D)$, any profile of $f(\theta|D)$ must be superficial. We here point out that, if it is admitted that the intractable $f(\theta|D)$ could result in such superficiality, why not the intuition for localization by enough data may imply the opposite. Hence, to invent and elaborate the language for $f(\theta|D)$ that can quantitatively carry the data-driven intuition should be a primary issue in the study of statistical inference.

We find out $f(\theta|D) = G_{\sigma_N}(\theta)$, which indicates that the localization of $f(\theta|D)$ near $\theta = \sigma_N$ is substantial. The upper bound of the width of $G_{\sigma_N}(\theta)$ characterizes and quantifies the localization. The objective existence of $G_{\sigma_N}(\theta)$ shows $f(\theta|D)$ is not such an entity that can be subjectively assigned in a belief. As $f(\theta|D)$ turns out to be a scientific entity, the frustrating situation by the denial of $f(\theta|D)$ in frequentism can be also resolved. Accepting the objectively specifiable central entity $f(\theta|D)$ will make the inference study natural and, for the first time, scientific. As a corollary, Eq. (1) immediately follows in the fact that any instance of $G_{\sigma_N}(\theta)$ is normalized and its width shrinks to zero in the $N \to \infty$ limit. In this way, the objective tractability of $f(\theta|D)$, we have established for exchangeable binary random variables in this article, shall shed a new light to the scientific study of statistical inference.

## Specified freedom as inference itself

The freedom of $G_{\sigma_N}(\theta)$ means various (restricted) possible forms of $f(\theta|D)$. Here, we emphasize that this freedom is far from an artifact, but rather a nature of inference. First of all, it is easy to see that there are infinitely many $f(\theta)$'s satisfying $P(D) = \int d\theta \prod_{i=1}^{N} \theta^{x_i}(1-\theta)^{1-x_i} f(\theta)$ for a given $P(D)$, de Finetti's representation theorem with which our derivation began. Then, multiple $f(\theta)$'s imply the associated freedom for $f(\theta|D)$ through $f(\theta) = \sum_D f(\theta|D)P(D)$. This freedom is manifested as the multiple possibilities for $G_{\sigma_N}(\theta)$ [see Eq. (4)]. Thus, to specify a single distribution as the inference result for a given observation is rather artificial. Based on Eqs. (3) and (4), statistical inference may be defined as the task of specifying the space of admissible distributions. This natural goal of statistical inference, which has been so far beyond mathematical description, is now achieved in the present work through the explicit representation of the distributional space $G_{\sigma_N}(\theta)$.

Interestingly, the freedom of $G_{\sigma_N}(\theta)$ does not allow to exclude the observation point $\theta = \sigma_N$ or to separate the distribution, i.e., any instance of $G_{\sigma_N}(\theta)$ is a distribution on one interval including $\theta = \sigma_N$. We remark that there has been an aspect called neutrosophic statistics [35], which comprehends the importance of degree of indeterminacy in statistical inference.

## Interval of predictive probability

An immediate application of $f(\theta|D)$ is $P(E|D) = \int d\theta P(E|\theta)f(\theta|D)$, the predictive probability of a future event $E$ after observation $D$. For $E$ of length $K$ with $k$ number of 1's, this predictive probability reads $P(E|D) = \int d\theta \theta^k(1-\theta)^{K-k} f(\theta|D) \equiv \langle \theta^k(1-\theta)^{K-k} \rangle_{f(\theta|D)}$. This average is basically a measure of the overlap between $\theta^k(1-\theta)^{K-k}$ and an instance of $f(\theta|D) = G_{\sigma_N}(\theta)$. Then, by the freedom of $G_{\sigma_N}(\theta)$, the values of $P(E|D)$ form the interval

$$\left( \inf\left\{ \langle \theta^k(1-\theta)^{K-k} \rangle_{G_{\sigma_N}(\theta)} \right\}, \sup\left\{ \langle \theta^k(1-\theta)^{K-k} \rangle_{G_{\sigma_N}(\theta)} \right\} \right). \cdots (5)$$

When $k/K \leq d_-$, for example, the interval becomes $\left( \langle \theta^K(1-\theta)^{K-k} \rangle_{g_{\sigma_N,d_+}(\theta)}, \langle \theta^K(1-\theta)^{K-k} \rangle_{g_{\sigma_N,d_-}(\theta)} \right)$. When $d_- < k/K < d_+$, the left end is trivially $\langle \theta^K(1-\theta)^{K-k} \rangle_{g_{\sigma_N,d_-}(\theta)}$ or $\langle \theta^K(1-\theta)^{K-k} \rangle_{g_{\sigma_N,d_+}(\theta)}$ while the right end is $\langle \theta^K(1-\theta)^{K-k} \rangle_{g_{\sigma_N,d}(\theta)}$ for some $d$. Similarly, other predictive probabilities for different $E$'s form their own intervals. The notion of the intervals of predictive probabilities is a natural consequence of the freedom of $G_{\sigma_N}(\theta)$. Therefore, the *predictive-probability interval* provides an objective language for prediction.

## From belief to science

In Bayesianism, the inference is $f(\theta|D) \propto P(D|\theta)f(\theta)$ for conditional probability $P(D|\theta)$ and belief $f(\theta)$ [1]. Since $f(\theta)$ is inherited, $f(\theta|D)$ becomes a belief. Even any $f(\theta|D)$ is possible in belief $f(\theta)$. In this situation, the actual step of inference is to choose a belief $f(\theta)$. This way, $f(\theta|D)$ is not a scientific entity in Bayesianism. However, in the consideration of the significance of statistical inference in science, we did not give up discerning $f(\theta|D)$ from belief, and have accomplished this goal by finding a new way where no belief on $f(\theta)$ is referred to. We did not accept such aspect that regards $P(D|\theta)$ as a belief too, which completely rules out a room to think of science.

We have discovered that the statistical inference is given by the Green's functions of the differential equation in Eq. (2), i.e., $f(\theta|D) = G_{\sigma_N}(\theta)$. As demonstrated above, no belief on $f(\theta)$ is referred to in deriving Eq. (2), in finding $f(\theta|D) = G_{\sigma_N}(\theta)$, and in obtaining $G_{\sigma_N}(\theta)$. Instead, the algebraic relations regarding $f(\theta)$ are used. Note the fact that a statement $f(\theta)$ is a belief does not mean a statement including $f(\theta)$ is a belief. For example, Bayes' theorem, a statement regarding four probabilities including $f(\theta)$, is not a belief but an identity. The coherency with Bayesianism of any inferred distribution is to be explained by the non-trivial homogeneous solutions that are extraneous. The rigorous procedure from de Finetti's representation theorem to the Green's function space, we have established in this work, shows a paradigm shift of statistical inference from belief to science. It is the shift from the belief of everything to the (philosophy of) science equipped with explicit computations free from a belief. Our study is based on the exchangeable binary random variables yet. Various generalizations such as beyond binary case or to higher order correction remain as future works.

## Statistical inference, a bottom line of science

Science pursues objectivity by inductive reasoning from observation data [36]. This objectivity is taken for granted while its concrete description may remain open. This normative issue of science was mathematically raised by Bayes in the late 18th century, and he arrived at the apparent subjectivity of statistical inference [1]. According to this subjectivity, science based on the inference from data should be a belief though this interpretation has been overlooked in the unprecedented success of science. To continue the success, science shall face the demand of the times, the scientific guide to the inference systems for big data expected or promoted in the name of artificial intelligence or machine learning. This basically asks what a bottom line of science, the inference from data, is. It is thus time to examine what the objectivity from data really means, and the rigorous procedure and result we have discovered in this work is an answer. Our finding will be a breakthrough in scientific statistical inference and science itself.

# Methods

## I. Sequence-generating operators

Consider the binary sequence $(x_1, x_2,.., x_N)$ for $x_i = 0$ or 1 with $\sum_i x_i = n$. Let $|N,n\rangle$ be a state notation for sequence length $N$ and the number of 1's therein $n$. We consider a linear operator $\hat{A}$, whose operation is given as [37]

$$\hat{A}|N,n\rangle = a_{N,n}|N+1, n+1\rangle \cdots (6)$$

for $0 < a_{N,n} < 1$, where $|N+1, n+1\rangle$ refers to $(x_1, x_2, .., x_N, 1)$. The linearity of $\hat{A}$ means that $\hat{A}(c|N,n\rangle + c'|N',n'\rangle) = c\hat{A}|N,n\rangle + c'\hat{A}|N',n'\rangle$ for constants $c$ and $c'$. The complementary operation to $\hat{A}$ is provided by

$$\hat{B}|N,n\rangle = b_{N,n}|N+1, n\rangle \cdots (7)$$

with $b_{N,n} = 1 - a_{N,n}$, where $|N+1, n\rangle$ refers to $(x_1, x_2, .., x_N, 0)$.

Beginning with the initial null state $|0,0\rangle$, upon a repeated application of $\hat{A}$ or $\hat{B}$ on $|0,0\rangle$, one may write the resulting state as

$$\left(\prod_{i=1}^{N} \hat{A}^{x_i} \hat{B}^{1-x_i}\right)|0,0\rangle = \left(\prod_{i=1}^{N} a_{i-1,\sum_{j=1}^{i-1} x_j}^{x_i} b_{i-1,\sum_{j=1}^{i-1} x_j}^{1-x_i}\right)|N,n\rangle$$
$$\equiv Q(x_1, \dots, x_N)|N,n\rangle \cdots (8)$$

for $x_i = 0$ or 1. By construction, it follows that $Q(x_1, x_2, \dots, x_N) > 0$ and $\sum_{x_1, x_2, \dots, x_N} Q(x_1, x_2, \dots, x_N) = 1$. This unit sum is attributed to the complementary property between $\hat{A}$ and $\hat{B}$ as readable in

$$\sum_{x_1, x_2, \dots, x_N} Q(x_1, x_2, \dots, x_N) = \sum_{x_1, x_2, \dots, x_{N-1}} \left(a_{N-1, \sum_{j=1}^{N-1} x_j} Q(x_1, x_2, \dots, x_{N-1})\right.$$
$$\left. + b_{N-1, \sum_{j=1}^{N-1} x_j} Q(x_1, x_2, \dots, x_{N-1})\right)$$
$$= \sum_{x_1, x_2, \dots, x_{N-1}} Q(x_1, x_2, \dots, x_{N-1}) = \cdots = \sum_{x_1} Q(x_1)$$
$$= a_{0,0} + b_{0,0} = 1. \cdots (9)$$

Therefore, $Q(x_1, x_2, \dots, x_N)$ can be interpreted as a probability measure when $x_i$ are binary random variables.

As a preliminary to our main task, we first show that $Q(x_1, x_2, \ldots, x_N)$ can play the role of the probability of sequence $(x_1, x_2, \ldots, x_N)$ that is composed of exchangeable binary random variables $x_i$.

## II. Equivalence between commutativity and exchangeability

Let $P(x_1, x_2, \ldots, x_N)$ be the probability for exchangeable binary random $x_i$. De Finetti's representation theorem [26] states that $P(x_1, x_2, \ldots, x_N) \equiv P_{N,n} = \int_0^1 d\theta\, \theta^n (1-\theta)^{N-n} f(\theta)$ for a mixture distribution $f(\theta)$. We will show that $\hat{A}$ and $\hat{B}$ with

$$a_{N,n} = \frac{P(x_1, x_2, \ldots, x_N, 1)}{P(x_1, x_2, \ldots, x_N)} = \frac{P_{N+1,n+1}}{P_{N,n}} \cdots (10)$$

commute with each other, that is, $\hat{A}\hat{B} = \hat{B}\hat{A}$ and that $Q(x_1, x_2, \ldots, x_N)$ generated by these $\hat{A}$ and $\hat{B}$ is no more than $P(x_1, x_2, \ldots, x_N)$.

The concrete meaning of $\hat{A}\hat{B} = \hat{B}\hat{A}$ in the $|N, n\rangle$-state space is $\hat{A}\hat{B}|N, n\rangle = \hat{B}\hat{A}|N, n\rangle$ for all $|N, n\rangle$. Hence, from Eqs. (6) and (7), it follows that

$$a_{N+1,n}(1 - a_{N,n}) = (1 - a_{N+1,n+1})a_{N,n} \cdots (11)$$

Our first interest is whether $a_{N,n}$ in Eq. (10) satisfies Eq. (11). This is straightforward by plugging Eq. (10) into Eq. (11) and then using $P_{N,n} = P_{N+1,n+1} + P_{N+1,n}$. Therefore, $\hat{A}$ and $\hat{B}$ with such $a_{N,n}$ in Eq. (10) commute with each other.

If $\hat{A}\hat{B} = \hat{B}\hat{A}$, then $\hat{A}$ and $\hat{B}$ can be arbitrarily shuffled in the left-hand side of Eq. (8). If all $\hat{A}$s are first applied, it follows that $Q(x_1, x_2, \ldots, x_N) = \prod_{i=1}^n a_{i-1,i-1} \prod_{i=n+1}^N b_{i-1,n}$. Using Eq. (10) and $P_{N,n} = P_{N+1,n+1} + P_{N+1,n}$, one then writes

$$Q(x_1, x_2, \ldots, x_N) = \prod_{i=1}^n \frac{P_{i,i}}{P_{i-1,i-1}} \prod_{i=n+1}^N \frac{P_{i,n}}{P_{i-1,n}} \cdots (12)$$

to find $Q(x_1, x_2, \ldots, x_N) = P_{N,n} = P(x_1, x_2, \ldots, x_N)$ after some trivial cancellations. For example, Pólya's urn model [38] and its variants [37,39,40] known to generate exchangeable sequences correspond to their own sampling procedures each of which is specifiable with the commutative pair $\hat{A}$ and $\hat{B}$ of $a_{N,n} = (n + \alpha)/(N + \alpha + \beta)$ for model-dependent constants $\alpha$ and $\beta$..

Therefore, $P(x_1, x_2, ..., x_N)$ of exchangeable random variables can be written as a repeated applications of commuting $\hat{A}$ and $\hat{B}$ to $|0,0\rangle$. This property plays a crucial role in the derivation of the differential equation of which Green's functions turn out to be the major elements of the statistical inference.

## III. $1/N$-expansion of the Laplace transform of $\sigma_N$

We define a linear operator $\partial_{\hat{A}}$ whose action is $\partial_{\hat{A}} \hat{A}^m = m\hat{A}^{m-1}$ for integer $m$ and introduce the operator $\langle \cdot |$ acting on any linear combination of $|m, n\rangle$ to give $\langle \cdot | \sum_{m,n} c_{m,n} |m, n\rangle = \sum_{m,n} c_{m,n}$. Let $P(x_1, x_2, ..., x_N)$ be the probability for the exchangeable $(x_1, x_2, ...)$ of binary $x_i$. Using the technique of the generating function [41,42], for the $m$-th moment of the random variable $\sigma_N \equiv (\sum_{i=1}^{N} x_i)/N$, one can write with a pair of commuting $\hat{A}$ and $\hat{B}$ as

$$\langle \sigma_N^m \rangle = \sum_{\sigma_N} \sigma_N^m P_{\sigma_N}$$

$$= \sum_{\sigma_N} \sigma_N^m \sum_{x_1, x_2, ..., x_N} P(x_1, x_2, ..., x_N) \delta_{N\sigma_N, \sum_{i=1}^{N} x_i}$$

$$= \sum_{x_1, x_2, ..., x_N} \left( \frac{\sum_{i=1}^{N} x_i}{N} \right)^m P(x_1, x_2, ..., x_N)$$

$$= \frac{1}{N^m} \langle \cdot | \left( (\hat{A} \partial_{\hat{A}})^m (\hat{A} + \hat{B})^N \right) | 0,0 \rangle, \cdots (13)$$

where $\delta_{i,j}$ is the Kronecker delta.

We shall make use a few relations to materialize Eq. (13). First, using the commutation relation $[\partial_{\hat{A}}, \hat{A}] \equiv \partial_{\hat{A}} \hat{A} - \hat{A} \partial_{\hat{A}} = I$, we get

$$(\hat{A} \partial_{\hat{A}})^m = \sum_{i=1}^{m} \mu_{m,i} \hat{A}^i \partial_{\hat{A}}^i \cdots (14)$$

for proper $\mu_{m,i}$. We then write

$$\langle \cdot | \hat{A}^i \partial_{\hat{A}}^i (\hat{A}+\hat{B})^N | 0,0 \rangle = \left( H(N-i) \prod_{j=0}^{i-1}(N-j) \right) \langle \cdot | \hat{A}^i (\hat{A}+\hat{B})^{N-i} | 0,0 \rangle$$

$$= \left( H(N-i) \prod_{j=0}^{i-1}(N-j) \right) \langle \cdot | (\hat{A}+\hat{B})^{N-i} \hat{A}^i | 0,0 \rangle$$

$$= \left( H(N-i) \prod_{j=0}^{i-1}(N-j) \right) \langle \cdot | (\hat{A}+\hat{B})^{N-i} \left( \prod_{j=0}^{i-1} a_{j,j} \right) | i,i \rangle$$

$$= \left( H(N-i) \prod_{j=0}^{i-1}(N-j) \right) \Theta_i \langle \cdot | (\hat{A}+\hat{B})^{N-i} | i,i \rangle$$

$$= \left( H(N-i) \prod_{j=0}^{i-1}(N-j) \right) \Theta_i, \cdots (15)$$

where $H(x)$ is the Heaviside step function, and $\Theta_i \equiv \prod_{j=0}^{i-1} a_{j,j}$. For the second equality in Eq. (15), $\hat{A}\hat{B} = \hat{B}\hat{A}$ is invoked; for the last equality, we utilize the straightforward algebraic result $\langle \cdot | (\hat{A}+\hat{B})^{N-i} | i,i \rangle = 1$ for all $N \geq i \geq 0$, which can be understood as the probability conservation. Using Eq. (10), one knows

$$\Theta_i = \prod_{j=0}^{i-1} \frac{P_{j+1,j+1}}{P_{j,j}} = \int_0^1 d\theta \, \theta^i f(\theta) \equiv \langle \theta^i \rangle_f . \cdots (16)$$

For $\prod_{j=0}^{i-1}(N-j)$ in the last line of Eq. (15), we below use

$$\prod_{j=0}^{i-1}(N-j) = \sum_{j=0}^{i-1}(-1)^j v_{i,j} N^{i-j} , \cdots (17)$$

for proper $v_{i,j}$.

Using Eqs. (14), (15), and (17), we find that Eq. (13) leads to

$$\langle \sigma_N^m \rangle = \delta_{m,0} + \frac{1}{N^m} \sum_{i=1}^{m} \mu_{m,i} H(N-i) \Theta_i \sum_{j=0}^{i-1} (-1)^j v_{i,j} N^{i-j}$$

$$= \delta_{m,0} + \frac{1}{N^m} \sum_{k=1}^{m} N^k \sum_{i=1}^{m} \mu_{m,i} H(N-i) \Theta_i \sum_{j=0}^{i-1} \delta_{k,i-j} (-1)^j v_{i,j}$$

$$= \delta_{m,0} + \frac{1}{N^m} \sum_{k=1}^{m} N^k \sum_{i=k}^{m} \mu_{m,i} H(N-i) \Theta_i (-1)^{i-k} v_{i,i-k}$$

$$= \delta_{m,0} + \frac{1}{N^m} \sum_{k=1}^{\min(m,N)} N^k \sum_{i=k}^{\min(m,N)} \mu_{m,i} \Theta_i (-1)^{i-k} v_{i,i-k}$$

$$= \delta_{m,0} + \sum_{k=1}^{\min(m,N)} \frac{1}{N^{m-k}} \sum_{j=0}^{\min(m,N)-k} (-1)^j \mu_{m,k+j} v_{k+j,j} \Theta_{k+j}, \cdots (18)$$

where $\min(x,y)$ is not the larger one out of $x$ and $y$.

We now expand the Laplace transform of $P_{\sigma_N}$, $\langle e^{-\sigma_N} \rangle$, in powers of $1/N$. Using Eq. (18), we have that

$$\langle e^{-s\sigma_N} \rangle = \left\langle \sum_{m=0}^{\infty} \frac{(-s\sigma_N)^m}{m!} \right\rangle$$

$$= 1 + \sum_{m=1}^{\infty} \frac{(-s)^m}{m!} \sum_{k=1}^{\min(m,N)} \frac{1}{N^{m-k}} \sum_{j=0}^{\min(m,N)-k} (-1)^j \mu_{m,k+j} v_{k+j,j} \Theta_{k+j}$$

$$= 1 + \sum_{i=0}^{\infty} \frac{1}{N^i} \sum_{m=i+1}^{i+N} \sum_{k=1}^{\min(m,N)} \delta_{i,m-k} \frac{(-s)^m}{m!} \sum_{j=0}^{\min(m,N)-k} (-1)^j \mu_{m,k+j} v_{k+j,j} \Theta_{k+j}$$

$$= 1 + \sum_{i=0}^{\infty} \frac{1}{N^i} \sum_{m=i+1}^{i+N} \frac{(-s)^m}{m!} \sum_{j=0}^{\min(m,N)+i-m} (-1)^j \mu_{m,m-i+j} v_{m-i+j,j} \Theta_{m-i+j}$$

$$= 1 + \sum_{i=0}^{\infty} \frac{1}{N^i} \sum_{k=1}^{N} \frac{(-s)^{i+k}}{(i+k)!} \sum_{j=0}^{\min(i+k,N)-k} (-1)^j \mu_{i+k,k+j} v_{k+j,j} \Theta_{k+j}$$

$$\equiv \sum_{i=0}^{\infty} \frac{1}{N^i} \Gamma_i(N,s). \cdots (19)$$

As the next step, we shall argue that $\lim_{N \to \infty} |\Gamma_i(N,s)| < \infty$ for all $i$.

## IV. $\lim_{N \to \infty} |\Gamma_i(N,s)| < \infty$ for all $i$

For a sequence composed of independent and identically distributed (IID) binary random variables $x_i$ of 1 or 0 with probability $p$ for $x_i = 1$, it is well known that [41,42]

$$\langle e^{-s\sigma_N} \rangle_{\text{IID}} = \left(1 - p + pe^{-s/N}\right)^N \cdots (20)$$

As an analytic continuation of Eq. (20), we consider a function $g(y,s) = (1 - p + pe^{-sy})^{1/y}$ with $g(0,s) = e^{-sp}$ for a continuous variable $y$. From the Taylor expansion of $g(y,s)$ around $y = 0$ for a fixed $s$, we find $|g^{(i)}(0,s)| < \infty$ where $g^{(i)}(0,s)$ is the $i$-th derivative of $g(y,s)$ at $y = 0$. Since IID sequence is exchangeable, Eq. (19) also holds for the IID case. Thus from $\langle e^{-s\sigma_N} \rangle_{\text{IID}} = \sum_{i=0}^{\infty}(1/N^i)\Gamma_i(N,s)_{\text{IID}}$, it follows that $\Gamma_i(\infty,s)_{\text{IID}} = g^{(i)}(0,s)$. One then knows $|\Gamma_i(\infty,s)_{\text{IID}}| < \infty$.

The right-hand side of Eq. (20) can be expanded into

$$\left(1 + p\sum_{n=1}^{\infty}\frac{(-s/N)^n}{n!}\right)^N = \sum_{i=0}^{\infty}\frac{1}{N^i}\sum_{j=0}^{N}\gamma_{i,j}(s)p^j \cdots (21)$$

Consequently, one may deduce that $\Gamma_i(\infty,s)_{\text{IID}} = \sum_{j=0}^{\infty}\gamma_{i,j}(s)p^j$ is convergent for all $p$ and $s$. Comparing it with Eq. (19), one has $\Gamma_i(\infty,s) = \sum_{j=0}^{\infty}\gamma_{i,j}(s)\Theta_j$. Then, for all $i$, it holds that

$$|\Gamma_i(\infty,s)| = \left|\sum_{j=0}^{\infty}\gamma_{i,j}(s)\Theta_j\right| \leq \sum_{j=0}^{\infty}\gamma_{i,j}(-|s|)\Theta_j \leq \sum_{j=0}^{\infty}\gamma_{i,j}(-|s|) < \infty \cdots (22)$$

based on the fact that $0 \leq \Theta_j = \langle \theta^j \rangle_f \leq 1$ [see Eq. (16)]. The last inequality of Eq. (22) comes from the IID case with $p = 1$.

## V. Differential equation with leading correction

That $\mu_{m,m} = \nu_{m,0} = 1$ is trivial, and $\mu_{m,m-1} = \nu_{m,1} = m(m-1)/2$ is easy to know. Using these, one may rewrite Eq. (19) as

$$\langle e^{-s\sigma_N} \rangle = \sum_{m=0}^{N}\frac{\langle(-s\theta)^m\rangle_f}{m!}$$

$$+ \frac{1}{2N}\sum_{m=0}^{N-1}\frac{(-s)\langle(-s\theta)^{m+1}\rangle_f - \langle(-s\theta)^{m+2}\rangle_f}{m!} + \mathcal{O}(1/N^2). \cdots (23)$$

We here can extend the upper bounds of the summations to $\infty$ whilst the $\mathcal{O}(1/N)$ terms are kept. Then, the right-hand side of Eq. (23) corresponds to $\mathcal{L}\{f(\theta)\}(s) + (s^2/2N)\mathcal{L}\{\theta(1-\theta)f(\theta)\}(s)$ in the ignorance of $\mathcal{O}(1/N^2)$ for large $N$, where $\mathcal{L}\{f(\theta)\}(s)$ is the Laplace transform [3] of $f(\theta)$ to $s$-domain. By applying the inverse Laplace transform on both sides of Eq. (23) while properly ignoring the higher order terms, we obtain

$$\sum_{\sigma_N} P_{\sigma_N} \delta(\theta - \sigma_N) = \left(1 + \frac{1}{2N}\frac{d^2}{d\theta^2}\theta(1-\theta)\right)f(\theta)$$

$$= \frac{1}{2N}\left(\theta(1-\theta)\frac{d^2}{d\theta^2} + (2-4\theta)\frac{d}{d\theta} - 2(1-N)\right)f(\theta)$$

$$\equiv \hat{L}_\theta f(\theta). \cdots (24)$$

Note that $\hat{L}_\theta$ is a hypergeometric differential operator [3], of which the homogeneous solution is given by a linear combination of hypergeometric functions $\mathcal{H}(\theta) = {}_2F_1(c_+, c_-, 2; \theta)$ for $c_\pm = (3 \pm \sqrt{8N+1})/2$ and $\mathcal{R}(\theta) = \mathcal{H}(1-\theta)$. We therefore arrive at Eq. (2).

## VI. Construction of $g_{\sigma_N,d}(\theta)$

Let $M_{\sigma_N}(\theta)$ represent the mathematical solutions of Eq. (24) for $P_{\sigma'_N} = \delta_{\sigma'_N, \sigma_N}$, that is, $\hat{L}_\theta M_{\sigma_N}(\theta) = \delta(\theta - \sigma_N)$. Then, these solutions can be formally written as $M_{\sigma_N}(\theta) = \sum_{i=0}^{N-1}(h_i \mathcal{H}(\theta) + r_i \mathcal{R}(\theta)) H(\theta - i/N) H((i+1)/N - \theta)$ with appropriate $h_i$ and $r_i$. As the simplest and meaningful ones out of $M_{\sigma_N}(\theta)$ in the viewpoint of probability distribution, we consider such $g_{\sigma_N,d}(\theta)$ that is positive in an interval including $\theta = \sigma_N$ and vanishes elsewhere, for which $\{h_i, r_i\}$ is additionally restricted. In the following, we demonstrate how these $h_i$ and $r_i$ are specified for the construction of $g_{\sigma_N,d}(\theta)$. We first consider the case where $1/N \leq \sigma_N \leq (1 - 1/N)$, and then the case of $\sigma_N = 0, 1$. We below shall often denote $N\sigma_N$ with $n$ for simplicity,

Firstly, $r_0 = 0$ is required because $\mathcal{R}(\theta)$ is not integrable on any interval including $\theta = 0$. Next, integration of Eq. (24) over the infinitesimal interval including $\theta = 0$ gives $h_0 \mathcal{H}(0^+) = 0$ in that there is no delta peak at $\theta = 0$ (the distribution magnitude is assumed to be zero for $\theta < 0$ without loss of generality). Then, with $r_0 = h_0 = 0$, we have the vanishing segment, $g_{\sigma_N,d}(\theta) = 0$ for $0 < \theta < 1/N$. As there is no delta peak at $\theta = i/N$ for $i \leq N\sigma_N - 1$ (note the current interest is $g_{\sigma_N,d}(\theta)$ whose inhomogeneity source is at $\theta = \sigma_N$ only), one may extend the vanishing profile up to $\theta = \sigma_N^-$ in the use of $h_i = r_i = 0$ while obeying Eq. (24). See the black bold line from $\theta = 0$ in Fig. 1(A).

Further extension of the vanishing profile to the slot labeled with $i = n = N\sigma_N$, that is, to the region of $\sigma_N < \theta < \sigma_N + 1/N$ is forbidden because there is unit heterogeneity at $\theta = \sigma_N$. The unit heterogeneity for Green's function and $h_i = r_i = 0$ for $i = 0,1,..,n-1$ are then accounted for in Eq. (24) to give

$$\left.\frac{d}{d\theta}\theta(1-\theta)\big(h_n\mathcal{H}(\theta) + r_n\mathcal{R}(\theta)\big)\right|_{\theta=\sigma_N^+} = 2N \cdots (25)$$

This shows that a non-trivial profile is required for $\sigma_N < \theta < \sigma_N + 1/N$, instead of vanishing one. The non-trivial profile appearing this way can be extended up to certain $\theta = (d/N)^-$ for an integer $d$ (by the assumed differentiability of the solution of Eq. (24) in each slots between adjacent delta functions, $d$ is an integer). This is done with $h_i = h_n$ and $r_i = r_n$ for $n + 1 \leq i \leq d - 1$. Note that such extension is compatible with no heterogeneity for $\theta > \sigma_N$. Beyond $\theta = d/N$, we assign the vanishing profile again till $\theta = 1$. See the black bold line for $\theta > d/N$ in Fig. 1(A). This is also consistent with no heterogeneity in $\theta > d/N$.

For the no heterogeneity at $\theta = d/N$, Eq. (24) requires

$$\left.\frac{d}{d\theta}\theta(1-\theta)\big(h_n\mathcal{H}(\theta) + r_n\mathcal{R}(\theta)\big)\right|_{\theta=(d/N)^-} = 0 \cdots (26)$$

Finally, the fact that $g_{\sigma_N,d}(\theta)$ must be a valid probability distribution dictates that

$$h_n\mathcal{H}(\theta) + r_n\mathcal{R}(\theta) \geq 0 \cdots (27)$$

for $\sigma_N < \theta < d/N$. This inequality imposes a maximum value $d_+ = d_+(\sigma_N)$ on $d$.

Solving Eqs. (25) and (26) for $h_n$ and $r_n$ in the inequality of Eq. (27), one obtains nontrivial part of $g_{\sigma_N,d}(\theta) = h\mathcal{H}(\theta) + r\mathcal{R}(\theta)$ for $\sigma_N < \theta < d/N$ with $h = h_n$ and $r = r_n$ depending on $d$. That is, each $d = n+1, n+2,..,d_+$ gives its own $g_{\sigma_N,d}(\theta)$. The blue bold curve in Fig. 1(A) represents an example obtained this way.

All the steps explained above are also possible in the backward direction from $\theta = 1$ to $\theta = 0$ while giving a non-trivial part on $d/N < \theta < \sigma_N$ with $d = d_-$, the minimum of $d$.

The bold curves drawn for $\theta < \sigma_N$ in Fig. 1(B) are the examples. When $n = 1$, only forward steps are possible with the trivial $d_- = 1$. Similarly, when $n = N - 1$, the case is backward only with $d_+ = N - 1$. The functions $g_{\sigma_N,d}(\theta)$'s constructed this way for both forward and backward directions form the building blocks in the construction of $G_{\sigma_N}(\theta)$.

When $\sigma_N = 0$, Eq. (24) with unit heterogeneity at $\theta = 0$ gives $G_0(\theta) = 2N\mathcal{H}(\theta)$ for $0 < \theta < 1/N$. One then finds $\int_0^{1/N} d\theta\, G_0(\theta) > 1$ by $\mathcal{H}'(0) = 1 - N$ and the convexity of $\mathcal{H}(\theta)$ in the integration interval; the former is a consequence of the well-known property of hypergeometric functions, $_2F_1'(a, b, c; x) = (ab/c)\, _2F_1(a + 1, b + 1, c + 1; x)$ [3], and the latter is backed up by numerical observation. Hence, $G_0(\theta)$ cannot meet the requirement for distribution. Similarly, when $\sigma_N = 1$, $G_1(\theta)$ is not a distribution because $\int_{1-1/N}^1 d\theta\, G_1(\theta) = \int_{1-1/N}^1 d\theta\, 2N\mathcal{R}(\theta) = \int_0^{1/N} d\theta\, G_0(\theta)$ by $\mathcal{R}(\theta) = \mathcal{H}(1 - \theta)$. Consequently, the associated building blocks like $g_{0,1}(\theta)$ or $g_{1,N-1}(\theta)$ do not exist.

# References

1. T. Bayes, R. Price, An essay towards solving a problem in the doctrine of chances, *Phil Trans* **53**, 370-418 (1763).

2. P. S. Laplace, *Essai philosophique sur les probabilités* (Courcier, Paris, 1814).

3. G. B. Arfken, H. J. Weber, *Mathematical Methods for Physicists* (Elsevier, Amsterdam, ed. 6, 2005).

4. J. Ladyman, *Understanding Philosophy of Science* (Routledge, Oxfordshire, 2002).

5. The Stanford Encyclopedia of Philosophy. URL: https://plato.stanford.edu/entries/confirmation.

6. Statistical inference. Encyclopedia of Mathematics. URL: http://encyclopediaofmath.org/index.php?title=Statistical_inference&oldid=37804.

7. J. M. Keynes, *A Treatise on Probability* (Macmillan & Co, London, 1921).

8. F. P. Ramsey (1931), *Truth and Probability (1926)*. The Foundations of Mathematics and other Logical Essays, Ch. VII, p.156-198, edited by R. B. Braithwaite, London: Kegan, Paul, Trench, Trubner & Co., New York: Harcourt, Brace and Company 1999 electronic edition.

9. J. M. Bernardo, A. F. M. Smith, *Bayesian Theory* (John Wiley & Sons, Chichester, 1994).

10. D. Corfield, J. Williamson, Eds., *Foundations of Bayesianism* (Springer, London, 2001).

11. E. T. Jaynes, *Probability Theory: The Logic of Science*, G. L. Bretthorst Ed. (Cambridge University Press, Cambridge, 2003).

12. R. A. Fisher, On the Mathematical Foundations of Theoretical Statistics, *Philosophical Transactions of the Royal Society of London, Series A* **222**, 309-368 (1922).

13. R. A. Fisher, *Statistical Methods for Research Workers* (Oliver & Boyd, Edinburgh, 1925)

14. J. Neyman and E. S. Pearson, On the Use and Interpretation of Certain Test Criteria for Purposes of Statistical Inference: Part I, *Biometrika* **20A**, 175-240 (1928)

15. J. Neyman, Outline of a Theory of Statistical Estimation Based on the Classical Theory of Probability, *Philosophical Transactions of the Royal Society, Series* A **236**, 333-380 (1937).

16. G. Gigerenzer, Z. Swijtink, T. Porter, L. Daston, J. Beatty, and L. Kruger, *The Empire of Chance: How Probability Changed Everyday Life* (Cambridge University Press, 1989).

17. L. Wilkinson, Statistical Methods in Psychology Journals, *American Psychologist* **54**, 594-604 (1999).

18. N. Colegrave, Confidence intervals are a more useful complement to nonsignificant tests than are power calculations, *Behavioral Ecology* **14**, 446-447 (2003).

19. J. H. McDonald, *Handbook of biological statistics*, 3rd ed. (Sparky House Publishing, Baltimore, 2014).

20. S. Greenland, S. J. Senn, K. J. Rothman, J. B. Carlin, C. Poole, S. N. Goodman, et. al., Statistical tests, P values, confidence intervals, and power: a guide to misinterpretations, *Eur. J. Epidemiol.* **31**, 337–350 (2016).

21. J. O. Berger, *The likelihood principle (Lecture notes-monograph series)* (Institute of Mathematical Statistics, 1984).

22. E.-J. Wagenmakers, A practical solution to the pervasive problems of p values, *Psychonomic Bulletin & Review* **14**, 779-804 (2007).

23. R. Hubbard and R. M. Lindsay, Why P Values Are Not a Useful Measure of Evidence in Statistical Significance Testing, *Theoretical Psychology* **18**, 69-88 (2008).

24. J. D. Perezgonzalez, Fisher, Neyman-Pearson or NHST? A tutorial for teaching data testing, *Frontiers in Psychology* **6**, 223-1-11 (2015).

25. R. D. Morey, R. Hoekstra, J. N. Rouder, M. D. Lee, E.-J. Wagenmakers, The fallacy of placing confidence in confidence intervals, *Psychon. Bull. Rev.* **23**, 103-123 (2016).

26. B. D. Finetti, La Prevision: Ses Lois Logiques, Ses Sources Subjectives, *Annales de l'Institut Henri Poincare* **7**, 1-68 (1937).

27. A. Bach, *Indistinguishable Classical Particles* (Springer, Berlin, 1997).

28. J. F. C. Kingman, Uses of exchangeability, *The Annals of Probability* **6**, 183-197 (1978).

29. D. V. Lindley, M. R. Novick, The role of exchangeability in inference, *The Annals of Statistics* **9**, 45-58 (1981).

30. O. Kallenberg, *Probabilistic Symmetries and Invariance Principles* (Springer, New York, 2000).

31. J. M. Stern, Symmetry invariance and ontology in physics and statistics, *Symmetry* **3**, 611-635 (2011).

32. J. K. Pritchard, M. Stephens, P. Donnelly, Inference of population structure using multilocus genotype data, *GENETICS* **155**, 945-59 (2000).

33. D. M. Blei, A. Y. Ng, M. I. Jordan, Latent Dirichlet allocation, *Journal of Machine Learning Research* **3**, 993-1022 (2003).

34. D. J. Hand, K. Yu, Idiot's bayes–not so stupid after all?, *International Statistical Review* **69**, 385-398 (2001).

35. For more information, see http://fs.unm.edu/NS/NeutrosophicStatistics.htm and references therein.

36. I. Newton, *Philosophiae Naturalis Principia Mathematica* (1687).

37. H. K. Lee, Y. W. Kim, Public opinion by a poll process: model study and Bayesian view, *J. Stat. Mech.: Theor. Exp.* **18**, 053402-11 (2018).

38. F. Eggenberger, G. Pólya, Über die statistik verketterer vorgänge, *J Appl Math Mech* **3**, 279-289 (1923).

39. N. Johnson, S. Kotz, *Urn Models and Their Applications* (Wiley, New York, 1977).

40. B. M. Hill, D. Lane, W. Sudderth, A strong law for some generalized urn processes, *Ann Probab* **8**, 214-226 (1980).

41. C. Gardiner, *Stochastic Methods: A Handbook for the Natural and Social Sciences* (Springer-Verlag, Berlin, ed. 4, 2009).

42. N. G. V. Kampen, *Stochastic Processes in Physics and Chemistry* (Elsevier, Amsterdam, ed. 3, 2007).